\documentclass[prc,superscriptaddress,showpacs,amsmath,amssymb,twocolumn]{revtex4-1}
\usepackage{amsmath, amsthm, amssymb, braket, revsymb}

\usepackage{amsfonts}

\usepackage{graphicx}
\usepackage{dcolumn}
\usepackage{bm}

\usepackage{color}

\newcommand{\be}{\begin{equation}}
\newcommand{\ee}{\end{equation}}

\newcommand{\ii}{\mathrm{i}}
\newcommand{\br}{\boldsymbol{r}}
\newcommand{\bsigma}{\boldsymbol{\sigma}}

\usepackage{ulem}

\begin{document}

\title{Suddenly shortened half-lives beyond $^{78}$Ni: $N=50$ magic number and high-energy non-unique first-forbidden transitions
}

\author{Kenichi Yoshida}
\affiliation{Department of Physics, Kyoto University, Kyoto, 606-8502, Japan}

\date{\today}

\begin{abstract}
{\noindent {\bf Background:} 
$\beta$-decay rates play a decisive role in understanding the nucleosynthesis of heavy elements and 
are governed by microscopic nuclear-structure information. 
A sudden shortening of the half-lives of Ni isotopes beyond $N=50$ was observed at the RIKEN-RIBF. 
This is considered due to the persistence of the neutron magic number $N=50$ in the very neutron-rich Ni isotopes. 
 \\
{\bf Purpose:} 
By systematically studying 
the $\beta$-decay rates and strength distributions in the neutron-rich Ni isotopes around $N=50$, 
I try to understand the microscopic mechanism for the observed sudden shortening of the half-lives.
\\
{\bf Methods:} 
The $\beta$-strength distributions in the neutron-rich nuclei 
are described in the framework of nuclear density-functional theory. 
I employ the Skyrme energy-density functionals (EDF) in the Hartree-Fock-Bogoliubov calculation 
for the ground states and in the proton-neutron Quasiparticle Random-Phase Approximation (pnQRPA) for the transitions. 
Not only the allowed but the first-forbidden (FF) transitions are considered. 
\\
{\bf Results:} 
The experimentally observed sudden shortening of the half-lives beyond $N=50$ is reproduced well 
by the calculations employing the Skyrme SkM* and SLy4 functionals 
in contrast to the monotonic shortening predicted in the preceding calculation using the SkO' functional. 
\\
{\bf Conclusions:} 
The sudden shortening of the half-lives beyond $N=50$ in the neutron-rich Ni isotopes 
is due to the shell gap at $N=50$ and cooperatively with the high-energy transitions 
to the low-lying $0^-$ and $1^-$ states in the daughter nuclei. 
The onset of FF transitions pointed out around $N=82$ and 126 is preserved in 
the lower-mass nuclei around $N=50$. 
This study suggests that 
needed is a microscopic calculation where the shell structure in neutron-rich nuclei 
and its associated effects on the FF transitions are selfconsistenly taken into account 
for predicting $\beta$-decay rates of exotic nuclei in unknown region. 
}
\end{abstract}

\maketitle

\section{Introduction}\label{intro}
Magic numbers are a key quantity to understand the quantum many-body systems 
such as atomic nuclei. 
This is because the single-particle motion in a mean-field potential and the associated shell structure 
are an important concept in nuclear structure. 
The local stability with the specific combinations of neutrons and protons is 
naturally understood with the help of the magic numbers. 
The canonical values were established in the systematic studies of nuclei 
along the $\beta$ stability line. 
In the recent studies of neutron-rich nuclei, however, it was found that 
some of the magic numbers disappear and 
the new ones show up instead~\cite{sor08}. 
Exploring the evolution of the shell structures and elucidating 
the underlying mechanism as functions of the neutron and proton numbers 
have been a fundamental research topic in the field of nuclear physics~\cite{ots18}. 

The neutron magic number $N=50$ around the very neutron-rich nucleus $^{78}$Ni $(N/Z \sim 1.8)$ 
has attracted a considerable interest 
not only in a view of nuclear structure but in a point of the $r$-process nucleosynthesis: 
$^{78}$Ni can be a doubly-magic nucleus and an important waiting point serving as a 
bottleneck in the synthesis of heavy elements~\cite{hos05}. 
Therefore, there have been a numerous number of 
experimental efforts on if the $N=50$ magic number survives 
in $^{78}$Ni~\cite{hak08,wel17, bar08, por12, zha04,pad05,van07,leb09,gad10,shi16,aoi10,tan14,xu14}. 

Though single-particle energies in a spherical mean field 
are not a direct observable, 
two-nucleon and one-nucleon separation energies are often considered as 
a corresponding signature. 
A high-precision nuclear mass measurement at the IGISOL on 
the $Z=30-33$ nuclei with a mass number around $80-90$ 
revealed a reduction of the $N=50$ shell gap energy towards 
Ge ($Z=32$) and an increase at Ga $(Z=31)$~\cite{hak08}, 
where the experimental shell gap energy was defined by the difference of the two-neutron separation energies as
$\Delta_{n}(N)=S_{2n}(N)-S_{2n}(N+2)$.  
By comparing with some mean-field calculations and mass models, 
Hakala {\it et al}. in Ref.~\cite{hak08} obtained an indication of 
the persistent $N=50$ shell gap in $^{78}$Ni. 
The recent measurement on $^{79}$Cu at the CERN-ISOLDE 
revealed a reduction of the $N=48$ shell gap, suggesting indirectly 
the enhanced $N=50$ shell gap~\cite{wel17}.
Another mass measurement on the Zn $(Z=30)$ isotopes 
including $^{81}$Zn confirmed that the $N=50$ shell gap is maintained 
for the Zn isotopes by looking at the one-neutron separation energies, 
and supports the indication above~\cite{bar08}. 
Porquet and Sorlin pointed out that the linear fits to the 
one-neutron separation energies lead to a diminished $N=50$ shell 
gap energy in $^{78}$Ni, where the collective effect at $Z=32$ 
was carefully examined~\cite{por12}. 
On the other hand, 
the large-scale shell model calculation explaining these observations 
predicts a persistent shell closure in $^{78}$Ni~\cite{sie12}.

The low-lying quadrupole state is sensitive to the 
softness of a spherical nucleus against quadrupole deformation. 
Thus, the excitation energy of the first-excited $2^+$ state 
and the transition matrix element are 
also a possible signature of the rigidity at magic numbers. 
A systematic investigation on the low- and medium-spin states 
in the $N=50$ isotones with $Z=32-37$ suggests a constant $N=50$ 
shell gap by comparing to the shell model calculation~\cite{zha04}. 
The lowering of transition probability $B(E2)$ in the Ge isotopes up to $^{82}$Ge 
also indicates that the $N=50$ shell closure remains in the neutron-rich nuclei~\cite{pad05,leb09,gad10}. 
The higher energy of the $2^+_1$ state and the lower $B(E2)$ value 
measured in $^{80}$Zn~\cite{van07}  
and the recently observed $4^+_1$ state~\cite{shi16} 
further confirmed the persistent $N=50$ shell closure in $^{78}$Ni. 
However, looking at the Ni isotopes, 
the $E(2_1^+)$ value decreases up to $N=48$ 
and the $B(E2)$ value increases up to $N=46$~\cite{aoi10}. 
Thus, the `direct' measurement of the $E(2_1^+)$ and $B(E2)$ in 
$^{78}$Ni is strongly desired~\cite{tan14}, while 
the recent theoretical calculations predict the doubly-closed structure 
in $^{78}$Ni~\cite{hag16, now16}.

Even though the production yields are small 
and the experiments on the mass or the low-lying quadrupole state 
are difficult to perform, 
the measurement of $\beta$-decay half-lives $T_{1/2}$ is feasible. 
Thus, there have been some attempts to study 
the structure of very neutron-rich nuclei via half-lives. 
The $\beta$-decay $T_{1/2}$ were discussed in terms of 
the shell effects, in particular the nuclear deformation effect, 
in the S and Cl nuclei with $N=28$~\cite{sor93} 
and in the $N \simeq 40$ nuclei around $^{64}$Cr~\cite{dau11}. 
In these works, the measured $T_{1/2}$ were compared with 
the results of model calculations 
such as the FRDM-QRPA~\cite{mol03} or its earlier version. 
The authors in Refs.~\cite{sor93, dau11} discussed 
how much of the deformation parameter is reasonable to explain the observation, 
and pointed out the shortcoming of the QRPA in which the deformation parameters 
for the mother and daughter nuclei are the same. 

Xu {\it et al}. carried out a systematic measurement of the $\beta$-decay 
$T_{1/2}$ of 20 neutron-rich nuclei in the $^{78}$Ni region at the RIKEN-RIBF~\cite{xu14}. 
They found a sudden shortening of $T_{1/2}$ in the Ni isotopes beyond $N=50$. 
When the $\beta$-decay $Q_\beta$ value is sufficiently high, 
$T_{1/2}$ is well approximated by a fifth-power dependence on $Q_\beta$. 
Experimentally, the monotonic shortening of $\log_{10}T_{1/2}$ of the Ni isotopes 
as a function of the neutron number was observed below $N=50$, 
indicating that the $Q_\beta$ value gradually increases in the isotopic chain below $^{78}$Ni. 
It was argued that the sudden shortening of $T_{1/2}$ beyond $N=50$ is due to the 
sudden increase in $Q_\beta$ because the nuclear deformation is not expected to occur: 
A large shell gap at $N=50$ gives us a dramatic increase in $Q_\beta$.  
As mentioned in Ref.~\cite{xu14}, the neutrons outside the $N=50$ gap may 
have a contribution to the $\beta$-decay via the forbidden transitions. 
However,  Borzov predicted that the first-forbidden (FF) transitions play only a minor role in 
the half-life of $^{79,80}$Ni, while it is important for describing the $\beta$-decay properties 
around $N \simeq 82$ and 126, 
and the $\beta$-delayed neutron emission probability around $^{78}$Ni~\cite{bor03}. 
Therefore, not only the shell gap but also the details of the nuclear wave functions 
are needed to investigate for understanding the origin of the observed 
sudden shortening of $T_{1/2}$ in the Ni isotopes. 

In the present article, I study systematically the $\beta$-decay $T_{1/2}$ 
in the neutron-rich Ni isotopes around $^{78}$Ni. 
Then, I try to understand the microscopic mechanism for the observed 
sudden shortening of the half-lives.
To this end, 
the $\beta$-strength distributions in the neutron-rich nuclei are described 
in a microscopic framework of nuclear density-functional theory. 
Here, not only the allowed but the FF transitions are considered on the same footing.

This article is organized in the following way: 
The theoretical frameworks for describing the ground state and 
the nuclear matrix elements needed for the $\beta$-decay rates are given in Sec.~\ref{theory}. 
And details of the numerical calculation are also given. 
However, a part of the details on the matrix elements needed for the FF transitions 
are given separately in Appendix.
Section~\ref{result} is devoted to the numerical results and discussion 
based on the microscopic calculation. 
Then, summary is given in Sec.~\ref{summary}.

\section{Theoretical framework}\label{theory}
\subsection{HFB and pnQRPA for the nuclear matrix elements}

In a framework of the nuclear energy-density functional (EDF) method I employed, 
the ground state of mother (target) nucleus is described by solving the 
Hartree-Fock-Bogoliubov (HFB) equation~\cite{dob84, bul80}:
\begin{align}
\begin{pmatrix}
h^{q}(\boldsymbol{r} \sigma)-\lambda^{q} & \tilde{h}^{q}(\boldsymbol{r} \sigma) \\
\tilde{h}^{q}(\boldsymbol{r} \sigma) & -(h^{q}(\boldsymbol{r} \sigma)-\lambda^{q})
\end{pmatrix}
\begin{pmatrix}
\varphi^{q}_{1,\alpha}(\boldsymbol{r} \sigma) \\
\varphi^{q}_{2,\alpha}(\boldsymbol{r} \sigma)
\end{pmatrix} 
= E_{\alpha}
\begin{pmatrix}
\varphi^{q}_{1,\alpha}(\boldsymbol{r} \sigma) \\
\varphi^{q}_{2,\alpha}(\boldsymbol{r} \sigma)
\end{pmatrix}, \label{HFB_eq}
\end{align}
where 
the mean field and the pair field are given by the functional derivative of the EDF 
with respect to the density and the pair density, respectively. 
The superscript $q$ denotes 
$\nu$ (neutron, $t_z = 1/2$) or $\pi$ (proton, $t_z =-1/2$).

The excited states $| i \rangle$ of the daughter nucleus are described as 
a one-phonon excitation built on the ground state $|0\rangle$ of the mother (target) nucleus:
\begin{align}
| i \rangle &= \hat{\Gamma}^\dagger_i |0 \rangle, \\
\hat{\Gamma}^\dagger_i &= \sum_{\alpha \beta}\left\{
X_{\alpha \beta}^i \hat{a}^\dagger_{\alpha,\nu}\hat{a}^\dagger_{\beta, \pi}
-Y_{\alpha \beta}^i \hat{a}_{\bar{\beta},\pi}\hat{a}_{\bar{\alpha},\nu}\right\},
\end{align}
where $\hat{a}^\dagger_\nu (\hat{a}^\dagger_\pi)$ and  $\hat{a}_\nu (\hat{a}_\pi)$ are 
the neutron (proton) quasiparticle creation and annihilation operators. 
The phonon states, the amplitudes $X^i, Y^i$ and the vibrational frequency $\omega_i$, 
are obtained in the proton-neutron quasiparticle random-phase approximation (pnQRPA).

The local one-body operators for the nuclear matrix elements relevant to
the $\beta$-decay rates of the axial-vector and vector type transitions are written by 
\begin{align}
^A\!\hat{O}_{\pm} &= \dfrac{1}{2}\sum_{\sigma, \sigma^\prime}  \sum_{\tau, \tau^\prime} \int d \br 
\langle \sigma| ^A\!O(\br, \bsigma)|\sigma^\prime \rangle 
\langle \tau|\tau_{\pm} |\tau^\prime \rangle
\hat{\psi}^\dagger(\br \sigma \tau) \hat{\psi}(\br \sigma^\prime \tau^\prime), \\
^V\!\hat{O}_{\pm} &= \dfrac{1}{2}\sum_{\sigma, \sigma^\prime}  \sum_{\tau, \tau^\prime} \int d \br ^V\!O(\br)\delta_{\sigma \sigma^\prime} 
\langle \tau|\tau_{\pm} |\tau^\prime \rangle
\hat{\psi}^\dagger(\br \sigma \tau) \hat{\psi}(\br \sigma^\prime \tau^\prime),
\end{align} 
respectively, where $\bsigma=(\sigma_{-1}, \sigma_0, \sigma_{+1})$ 
denotes the spherical components of the Pauli spin matrices, and 
$\tau_{\pm}=(\tau_x \pm \ii \tau_y)$ are the isospin-ladder operators. 
The nucleon field operators $\hat{\psi}^\dagger$ and $\hat{\psi}$ are expressed 
in terms of the quasiparticle operators $\hat{a}^\dagger$ and $\hat{a}$ 
through the generalized Bogoliubov transformation. 
I consider the allowed and first-forbidden (FF) transitions in the present calculation. 
For the allowed transitions, the operators needed are just for $^A\!O(\br, \bsigma)=\bsigma$ and $^V\!O(\br)=1$, 
while for the FF transitions, one needs the operators for 
those containing $rY_1$ and those appearing due to 
the relativistic correction as summarized in Table~\ref{ope}. 
Here, the operators $^A\!\hat{O}^{J 1 K}$ and $^V\!\hat{O}^{11K}$ correspond 
to the charge-exchange rank-$J$ spin-dipole and dipole operators, respectively. 
The nuclear transition matrix elements $\langle i | \hat{O}| 0\rangle$ are evaluated 
in the standard quasi-boson approximation as $\langle \mathrm{HFB}| [\hat{\Gamma}_i, \hat{O}]|\mathrm{HFB}\rangle$.

\begin{table}[t]
\begin{center}
\caption{Summary of $^A\!O(\br,\bsigma)[^V\!O(\br)]$ in the 
operators $^{A(V)}\!\hat{O}^{J L K}$ needed for the matrix elements of the FF transitions, 
where $m_{\mathrm{n}}$ is the mass of the nucleon. 
The factor $\Theta_K$ arises from the transformation from the intrinsic to the laboratory frames of reference~\cite{boh75}: 
$\Theta_K=1$ and $\sqrt{2}$ for $K=0$ and $K \ne 0$, respectively. }
\label{ope}
\begin{tabular}{cccc}
\hline \hline
\noalign{\smallskip}
 ${}^A\!\hat{O}^{000}$ & ${}^{A}\!\hat{O}^{J 1 K}$ & ${}^{V}\!\hat{O}^{10K}$ & ${}^{V}\!\hat{O}^{11K}$ \\
\noalign{\smallskip}\hline\noalign{\smallskip}
 $\dfrac{\bsigma \cdot \nabla}{m_\mathrm{n}}$ & $\sqrt{\dfrac{4\pi}{3}}r[Y_1 \otimes \bsigma]^J_K\Theta_K$  &
$\dfrac{\nabla_K}{m_\mathrm{n}}\Theta_K$ & $\sqrt{\dfrac{4\pi}{3}}rY_{1K}\Theta_K$ \\
\noalign{\smallskip}
\hline \hline
\end{tabular}
\end{center}
\end{table}

\subsection{numerical calculations}
I solved the coordinate-space HFB equations 
in the cylindrical coordinates
$\boldsymbol{r}=(\rho,z,\phi)$ with a mesh size of
$\Delta\rho=\Delta z=0.6$ fm and a box
boundary condition at $(\rho_{\mathrm{max}},z_{\mathrm{max}})=(14.7, 14.4)$ fm.
The qp states were truncated according to the qp 
energy cutoff at 60 MeV, and 
the qp states up to the magnetic quantum number $\Omega=23/2$
with positive and negative parities were included. 

The two-body interactions for the pnQRPA equation were derived self-consistently from the EDF. 
I introduced the truncation for the two-quasiparticle (2qp) configurations in the QRPA calculation,
in terms of the 2qp-energy as 60 MeV. 
More details of the calculation scheme are given in Ref.~\cite{yos13}. 

For the normal (particle-hole) part of the EDF,
I employed mainly the SkM* functional~\cite{bar82} and secondarily the SLy4 functional~\cite{cha98}. 
For the pairing energy, I adopted the one in Ref.~\cite{yam09}
that depends on both
the isoscalar and isovector densities, 
in addition to the pairing density, with the parameters given in
Table~III of Ref.~\cite{yam09}. 
The same pairing EDF was employed for the $T=1$ pn-pairing interaction 
in the pnQRPA calculation, 
while the linear term in the isovector density was dropped.
The $T=0$ pairing interaction 
was also included in the present pnQRPA calculation, 
with the same strength as the $T=1$ pairing interaction. 
Though I did not optimize the pairing strengths, 
the characteristic isotopic dependence was reproduced well by  
the present work. 

\subsection{calculation of the $\beta$-decay rates}

The $\beta$-decay rate $\lambda_\beta$ and 
the partial half-life $t_{1/2}$ including the allowed and FF transitions 
can be calculated as~\cite{beh82, sch66},
\begin{align}
\dfrac{1}{t_{1/2}}&=\dfrac{\lambda_\beta}{\ln 2}=\dfrac{f}{D}, \\
f&=\int_1^{W_0} C(W)F(Z,W)p W (W_0-W)^2 dW,
\label{beta_rate}
\end{align}
where I used $D=6147$ s for the constant, and $C(W)$ 
is the shape factor containing the nuclear matrix elements as described below. 
The Fermi function $F(Z,W)$ in (\ref{beta_rate}) 
including the effect of the Coulomb distortion on the electron wave function 
is given by
\begin{equation}
F(Z,W)=2(1+\gamma)(2pR)^{-2(1-\gamma)}e^{\pi \nu}
\left|\dfrac{\Gamma(\gamma + \ii\nu)}{\Gamma(2\gamma +1)}\right|^2,
\end{equation}
where $\gamma=\sqrt{1-(\alpha Z)^2}$, $\nu=\alpha ZW/p$, $\alpha$ is the fine structure 
constant, and $R$ is the nuclear radius calculated as $1.2\times A^{1/3}$ fm. 
$W$ is the total energy of the electron, 
$W_0$ 
is the total energy available; 
\begin{align}
W_0 &= m_e c^2 + \lambda_\nu - \lambda_\pi + \Delta M_{n-H} - \omega_i \\
&=m_e c^2 -E_{\mathrm{T},i} + \Delta M_{n-H},
\end{align}
and $p=\sqrt{W^2-1}$ is the momentum. 
$\Delta M_{n-H}$=0.782 MeV is a mass difference between 
a neutron and a hydrogen atom, and $E_{\mathrm{T},i}$ the excitation energy 
with respect to the ground-state of the mother (target) nucleus, 
and $\lambda_{\pi(\nu)}$ the chemical potential of protons (neutrons).
I use natural units $\hbar=c=m_e=1$. 
The unit of length is the reduced electron Compton wavelength, 
$\lambdabar_e =386.16$ fm. 

The shape factor for the allowed transitions is given as
\begin{equation}
C_0=|\langle 1 \rangle|^2+
\lambda^2
|\langle \boldsymbol{\sigma} \rangle|^2,
\end{equation}
where $\langle \cdot \rangle$ denotes 
the nuclear transition matrix element 
between the ground and excited states $|i\rangle$ 
for the isospin lowering operator, and 
$\lambda=-(g_A/g_V ) = 1.2701(25)$ 
is the ratio of weak axial and vector coupling constants. 
I used the quenching factor for the allowed Gamow-Teller transitions, $q=(1/1.27)^2=0.62$ or equivalently $\lambda=1$, 
as commonly used in the pnQRPA framework~\cite{eng99, mol03,mus16, mar16,sar18}.
Since the shape factor $C_0$ is independent of the energy $W$, 
it is convenient to define the phase space factor, or so-called the integrated Fermi function $f_0$ as
\begin{align}
ft_{1/2} &= C_0 f_0 t_{1/2} = D, \\
f_0 &=\int_1^{W_0}F(Z,W)pW (W_0-W)^2 dW.
\end{align}

When the FF transitions take part in the $\beta$-decay process, 
I need to consider the explicit dependence of the shape factors 
on the electron energy as the details are given in Appendix~\ref{app_FF}.
To discuss the FF transition strengths distribution and compare with the allowed transitions 
relevant to the $\beta$-decay, 
it is useful to define the averaged shape factor as~\cite{beh82}
\begin{equation}
\overline{C(W)}=\dfrac{f}{f_0}
\end{equation}
so that the partial half-life is related to $\overline{C(W)}$ as $D/f_0 t_{1/2}$ 
similarly to the allowed transitions.

\section{Results and discussion}\label{result}

\begin{figure}[t]
\begin{center}
\includegraphics[scale=0.25]{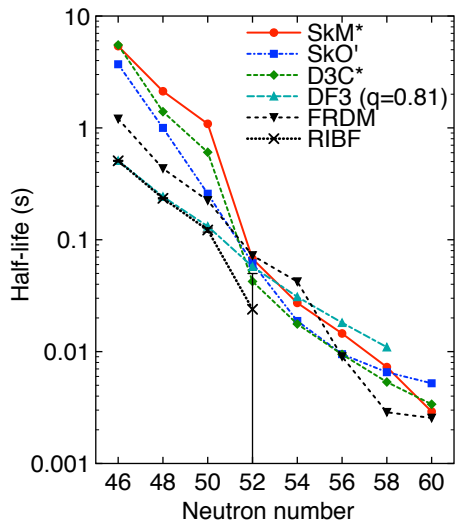}
\caption{Calculated $\beta$-decay half-lives of the neutron-rich Ni isotopes with use of the SkM* functional. 
Shown are together with the results in the preceding pnQRPA calculations~\cite{mol03, bor05, mus16, mar16} 
and the half-lives measured at the RIKEN-RIBF~\cite{xu14}.}
\label{beta_decay1}
\end{center}
\end{figure}

Figure~\ref{beta_decay1} shows the calculated $\beta$-decay half-lives of 
the neutron-rich even-$N$ Ni isotopes by using the SkM* functional. 
One clearly sees a sudden shortening of the half-lives beyond $N=50$. 
Up to $N=50$ and above $N=52$, the calculation shows a monotonic decrease in the half-lives. 
One can thus expect something singular to have happened between $N=50$ and 52. 
In what follows, I am going to discuss the mechanism for this 
sudden shortening of the half-lives beyond $N=50$. 

Before discussing the isotopic dependence of the calculated half-lives, 
I briefly mention the results obtained in the preceding pnQRPA calculations, 
which are also shown in Fig.~\ref{beta_decay1}.  
The calculation employing the relativistic functional D3C*~\cite{mar16} gives 
the similar result to that using the SkM* functional, while the other microscopic calculations 
shown here~\cite{mus16, bor05} produce the monotonic decrease. 
Taking a closer look at the result of the FRDM~\cite{mol03}, 
where the FF transitions are taken into account by the gross theory, one sees that 
the model predicts a sudden shortening beyond $N=54$. 
The calculations except that using DF3~\cite{bor05}, where the large quenching factor $q=0.81$ 
was used, overestimate the observed half-lives. 
Though it is beyond the scope of the 
present work, there is room for a further investigation on the 
beyond-RPA effects such as in a particle-vibration coupling scheme as discussed in Ref.~\cite{niu16}.

\begin{figure}[t]
\begin{center}
\includegraphics[scale=0.25]{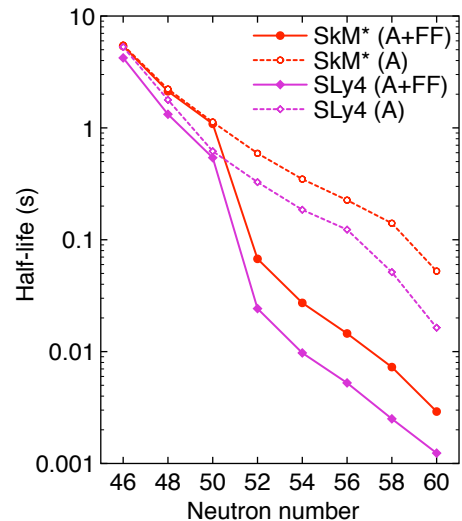}
\caption{Similar to Fig.~\ref{beta_decay1} but employing the SLy4 functional 
besides the SkM* functional. 
Shown are also the results obtained in the allowed approximation denoted by the open symbols.
 }
\label{beta_decay2}
\end{center}
\end{figure}

I show in Fig.~\ref{beta_decay2} 
the calculated $\beta$-decay half-lives obtained in the allowed (A) approximation and 
those where the FF transitions are also taken into account (A+FF). 
The calculations in the A approximation only produce the 
monotonic shortening, while those in the A+FF approximation 
give a sudden shortening at $N=52$. 
Not only the calculation using SkM* but that using the SLy4 functional 
produce the sudden shortening.
Thus, the FF transitions are necessary to explain the observed 
sudden shortening of the $\beta$-decay half-lives. 
The calculation scheme in Ref.~\cite{mus16} is analogous to the present one 
in the sense that the Skyrme-type EDF was employed for the HFB+pnQRPA 
calculation and the FF transitions were taken into account in a microscopic way. 
The only difference to the present calculation is that 
the SkO' functional~\cite{rei99} was used there. 
I am going to investigate the reason why the SkM*, SLy4, and D3C* functionals 
produce the sudden shortening, while SkO' only produces the monotonic 
one for the $\beta$-decay half-lives, 
then unravel the mechanism for the sudden shortening beyond $N=50$.

\begin{figure}[t]
\begin{center}
\includegraphics[scale=0.25]{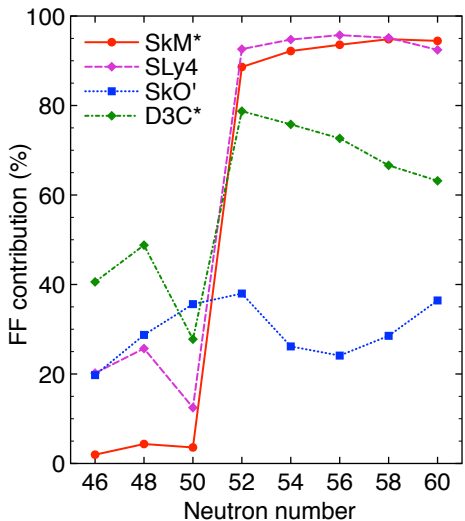}
\caption{FF contribution to the total $\beta$-decay rate computed using 
the Skyrme-type EDFs of SkM* and SLy4 in the present calculation, 
and SkO' in Ref.~\cite{mus16}, and the 
relativistic EDF of D3C* in Ref.~\cite{mar16}.}
\label{FF_contribution}
\end{center}
\end{figure}

Figure~\ref{FF_contribution} displays the FF contribution to 
the computed total $\beta$-decay rate. 
The calculations using the SkM*, SLy4, and D3C* functionals show that 
the FF transitions largely contribute to the $\beta$-decay rate 
at $N=52$ while the FF contribution is low at $N=50$. 
However, 
the calculation using SkO' only shows the monotonic increase 
in the FF contribution as the neutron number increases from 46 to 52, 
and the much less FF contribution than the others for $N \geq 54$. 
Therefore, the onset of the FF transitions beyond $N=50$ 
is a key to the understanding of the mechanism for the 
sudden shortening of the $\beta$-decay half-lives 
in the neutron-rich Ni isotopes. 

\begin{figure}[t]
\begin{center}
\includegraphics[scale=0.25]{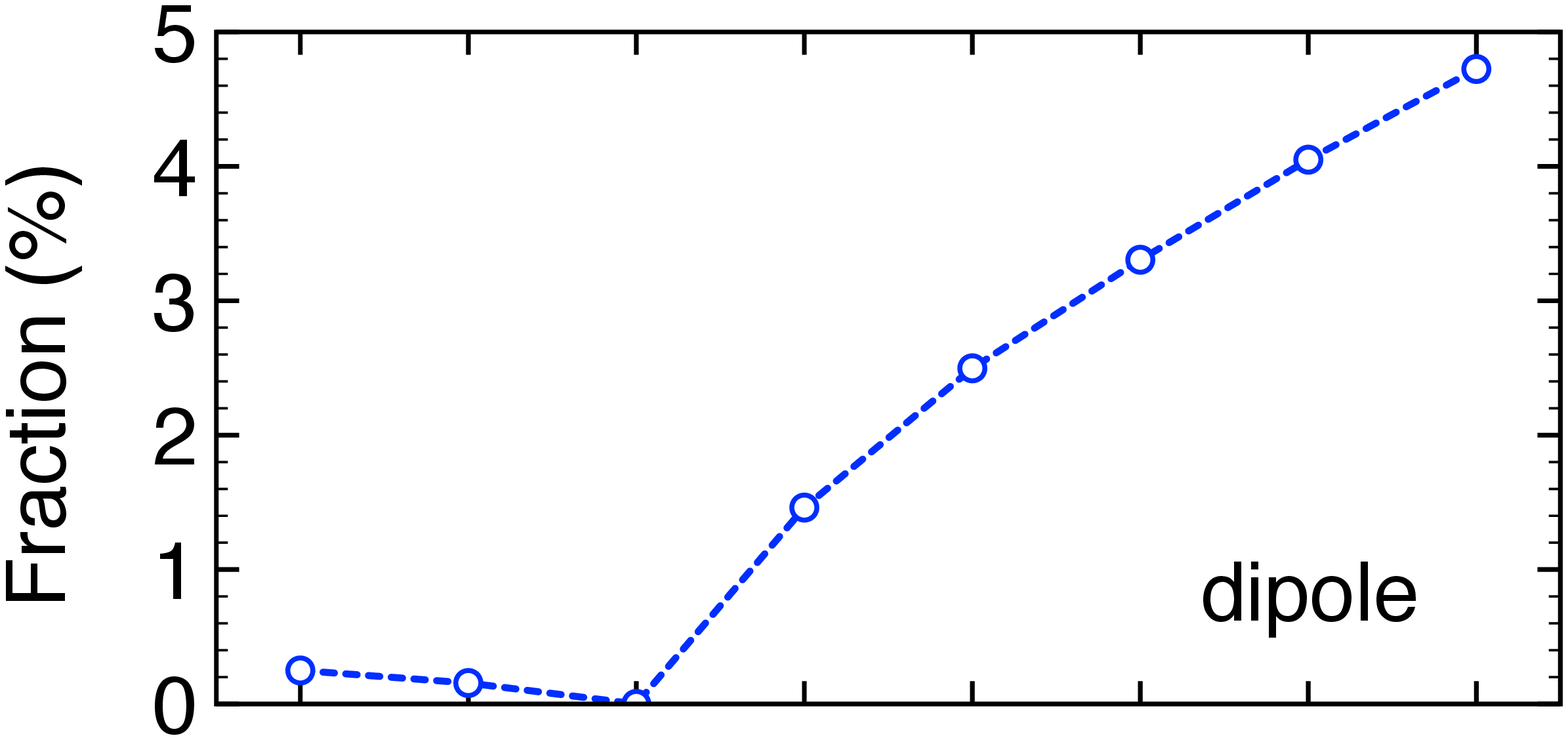}
\includegraphics[scale=0.25]{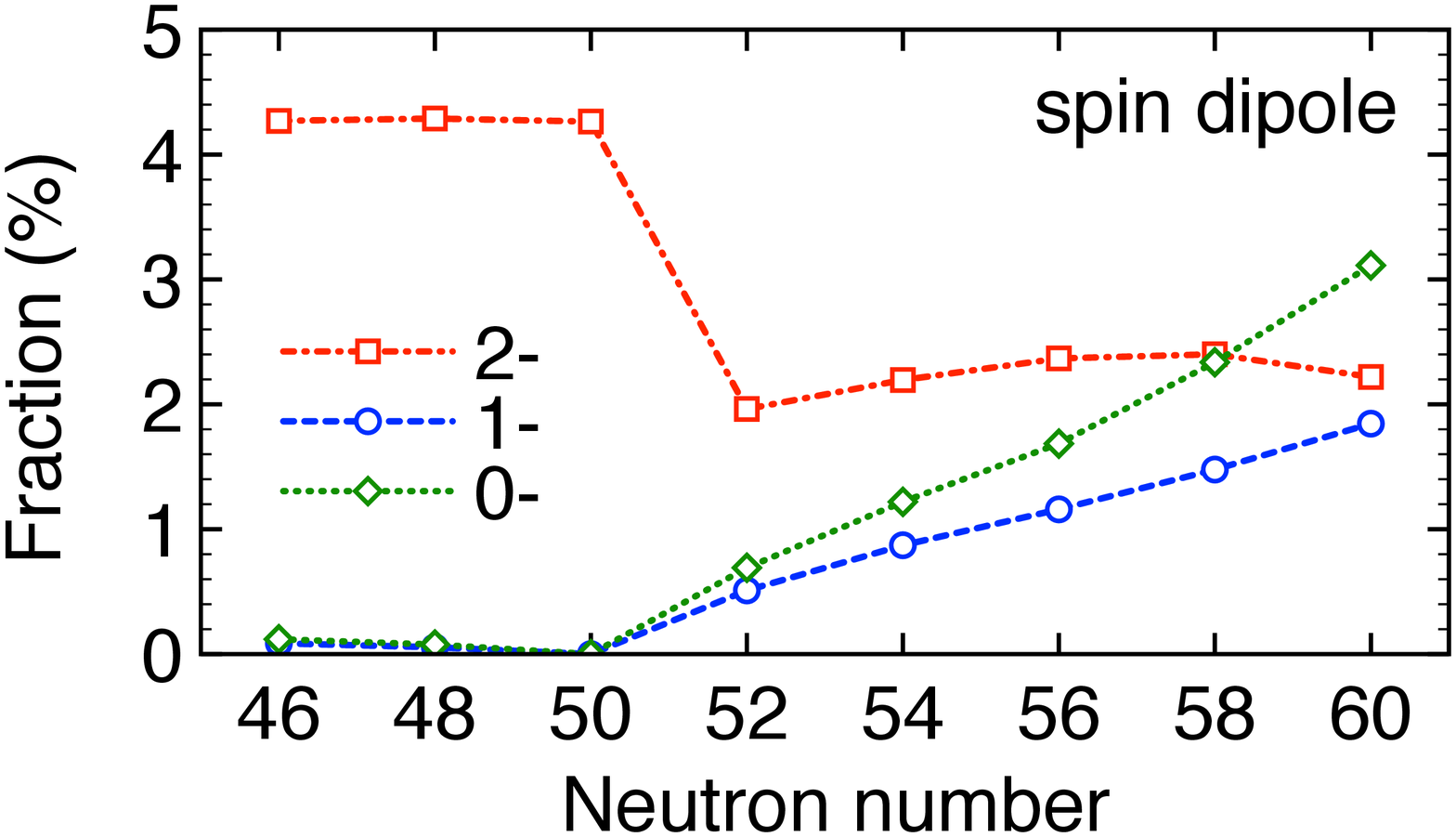}
\caption{Fraction of the summed strengths of the dipole (top) and the spin dipole (bottom) 
transitions in low energy below $E_{\mathrm{T}}<0$ calculated using the SkM* functional.
}
\label{strengths}
\end{center}
\end{figure}

\begin{figure*}[t]
\begin{center}
\includegraphics[scale=0.28]{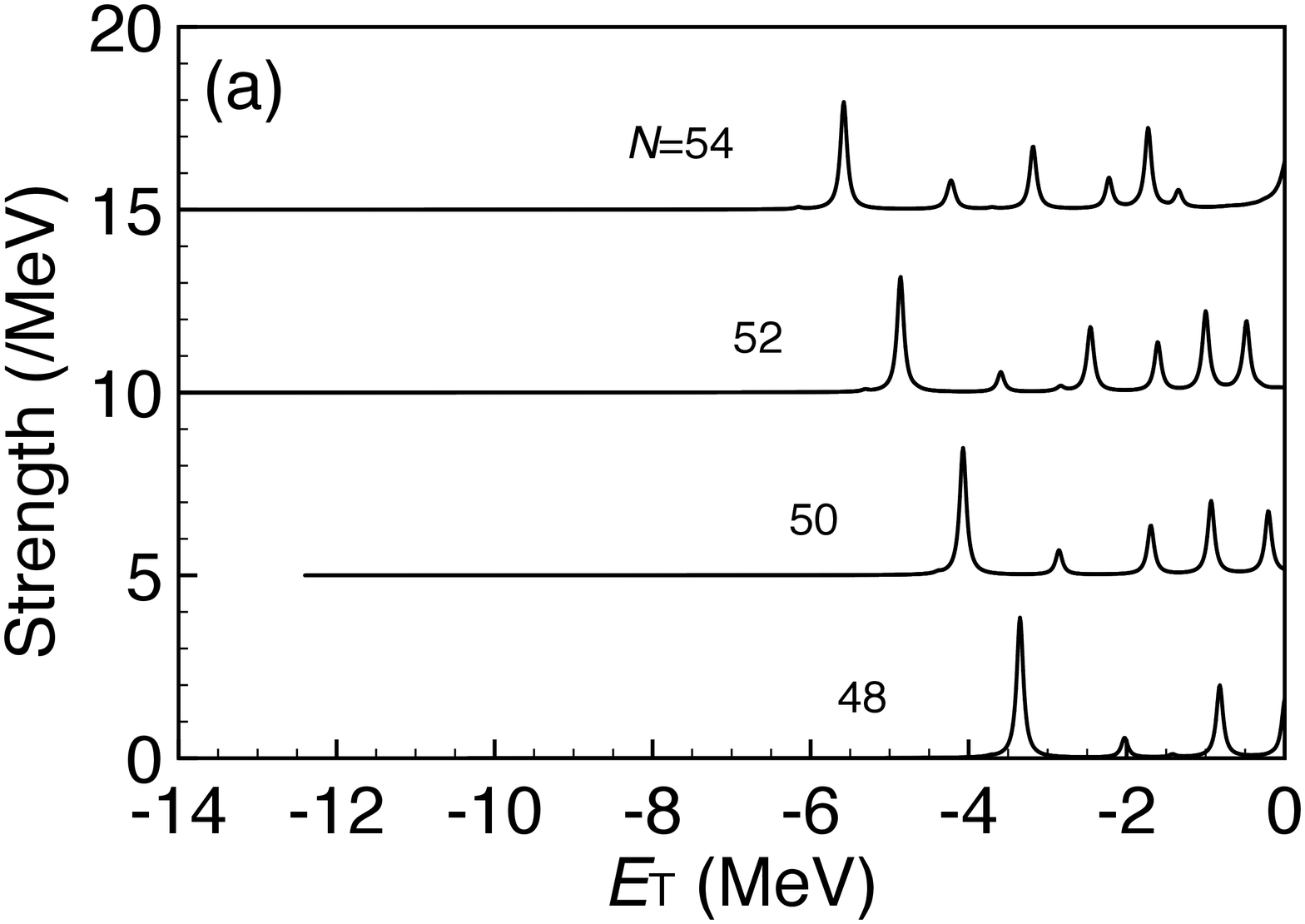}
\includegraphics[scale=0.28]{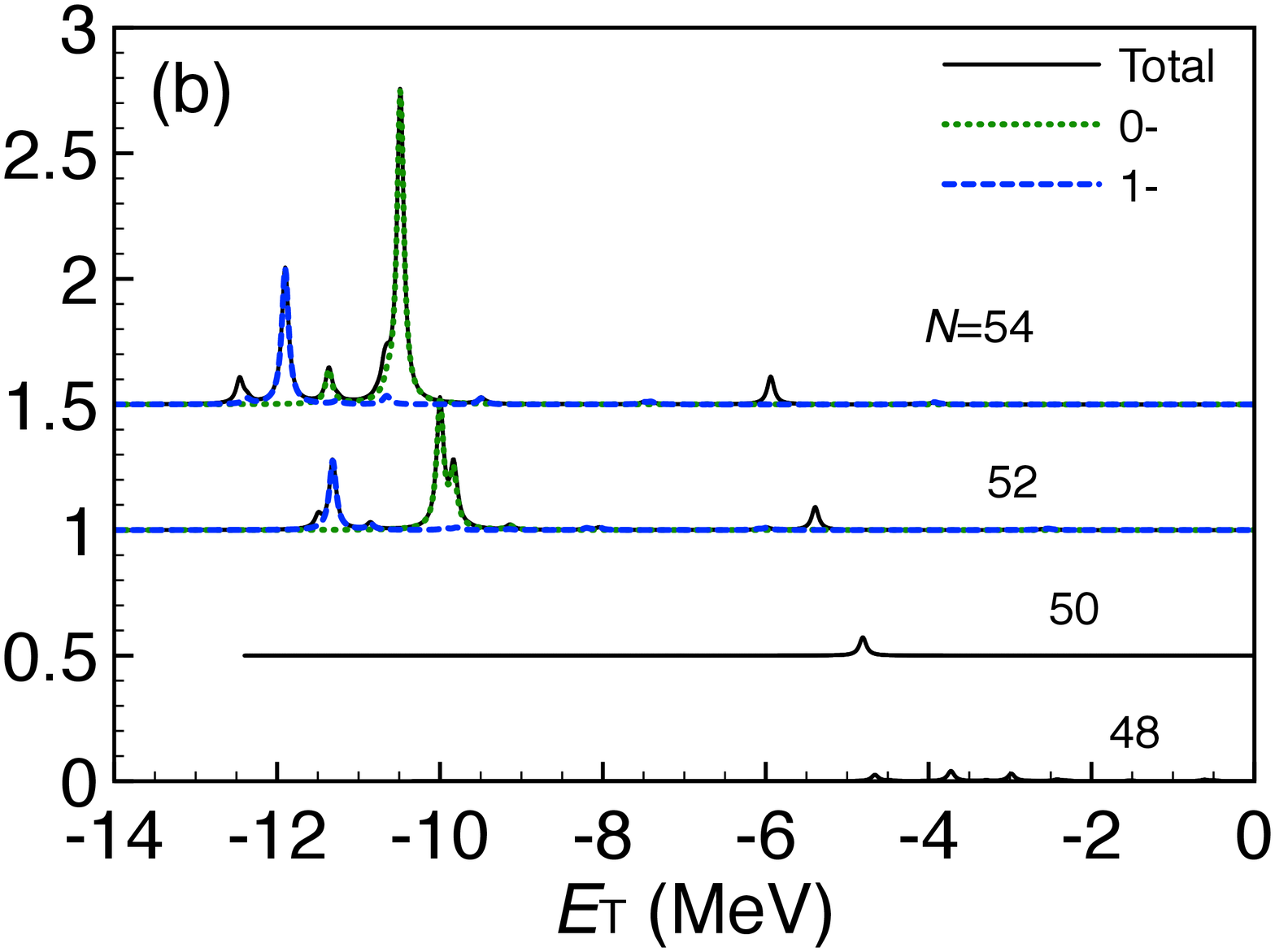}
\caption{Shape factors and the averaged shape factors for (a) the allowed and 
(b) the FF transitions from 
$^{76-82}$Ni calculated by using the SkM* functional.}
\label{beta_strength}
\end{center}
\end{figure*}

The onset of the FF transitions beyond $N=50$ 
is simply understood by the appearance 
of the low-energy $-1\hbar \omega_0$ excitation 
associated with the shell structure in neutron-rich nuclei~\cite{yos17}. 
In the light nuclei around $A\simeq 40$, the similar mechanism for the 
onset of the FF transitions beyond $N=28$ was also discussed~\cite{yos18}.
Figure~\ref{strengths} shows the fraction of the summed strengths of the dipole ($\Delta J^\pi=1^-$) 
and the spin dipole ($\Delta J^\pi=0^-, 1^-,$ and $2^-$) 
transitions in low energy below $E_{\mathrm{T}}<0$. 
The upper figure in Fig.~\ref{strengths} 
corresponds to Fig.~3(b) in Ref.~\cite{yos17}. 
As the neutrons start to occupy the $2d_{5/2}$ and $2d_{3/2}$ orbitals, 
the negative-parity $\nu d_{5/2}, d_{3/2} \to \pi p_{3/2}, p_{1/2}$ excitations are possible to occur. 
From this figure, one sees that the $0^-$ and $1^-$ excitations are 
crucial for the sudden decease in $T_{1/2}$ between $N=50$ and 52. 
The contribution of the $2^-$ excitation is anti-correlated with the decrease in $T_{1/2}$. 
Figure~\ref{beta_strength} displays 
the shape factors and the averaged shape factors for the allowed and 
the FF transitions from $^{76-82}$Ni. 
As can be seen in Fig.~\ref{beta_strength}(b), there show up several 
negative-parity states possessing an appreciable strength.

\begin{figure}[t]
\begin{center}
\includegraphics[scale=0.25]{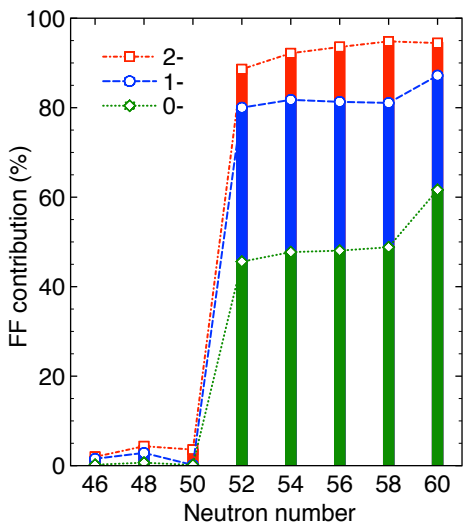}
\caption{Contributions of the different FF multipoles to the total $\beta$-decay rate 
computed using the SkM* functional.
}
\label{FF_SkM*}
\end{center}
\end{figure}

I am going to discuss further the microscopic origin of the sudden onset 
of the FF transitions beyond $N=50$. 
I show in Fig.~\ref{FF_SkM*} the contributions of the 
FF multipoles to the total $\beta$-decay rate obtained by 
using SkM*. 
One clearly sees that the $0^-$ and $1^-$ transitions are dominant 
beyond $N=50$. 
This can be understood by the location of the Fermi level of neutrons. 
Up to $N=50$, the $g_{9/2}$ is the only positive-parity occupied orbital. 
So, the $\nu g_{9/2} \to \pi f_{5/2}$ FF transition with $\Delta J^\pi=2^-$ 
is only possible to occur. 
When the Fermi level moves to the $d_{5/2}$ orbital, 
the partially occupied neutrons in the $d_{5/2}$ orbital can participate in 
the $\nu d_{5/2}\to \pi f_{5/2}$ excitation with $\Delta J^\pi=0^-$ and 
the $\nu d_{5/2}\to \pi p_{3/2}$ excitation with $\Delta J^\pi=1^-$ 
besides the $\nu d_{5/2}\to \pi p_{1/2}$ excitation with 
$\Delta J^\pi=2^-$ transition. 
One sees these excitations 
in the averaged shape factors for the FF transitions 
displayed in Fig.~\ref{beta_strength}(b). 
The low-lying prominent peaks in $^{80,82}$Ni correspond to the 
$\Delta J^\pi=0^-$ and $1^-$ excitations. Furthermore, 
in the relatively higher energy region $E_{\mathrm{T}}\sim -5$ MeV, 
one sees the $\nu g_{9/2} \to \pi f_{5/2}$ excitation.

One more important thing for the sudden onset of the FF transitions 
beyond $N=50$ is the magnitude of the relative energy 
between the FF and allowed transitions.
One sees in Fig.~\ref{beta_strength}(a) 
a prominent peak in the shape factor around $E_\mathrm{T}\sim -5$ MeV, 
which is above the FF transition by about 5 MeV. 
This is predominantly constructed by the $\nu p_{1/2} \to \pi p_{3/2}$ excitation with 
the RPA amplitude $X^2-Y^2$ being greater than 0.9. 
The transition strength for this allowed transition is 
much larger than that for the FF transitions.
Since the phase space factor $f_0$ roughly behaves as $W_0^5$, 
the FF transitions are able to overcome the contribution from the allowed transitions 
if they are high enough in energy, and are well apart from the allowed transitions. 
In the present case, the FF transitions gain a factor $\sim (10/5)^5=32$ from the phase space factor. 
Therefore, the contribution from the FF transitions at 
$N=52$ dominates that from the allowed transitions.

\begin{figure}[t]
\begin{center}
\includegraphics[scale=0.2]{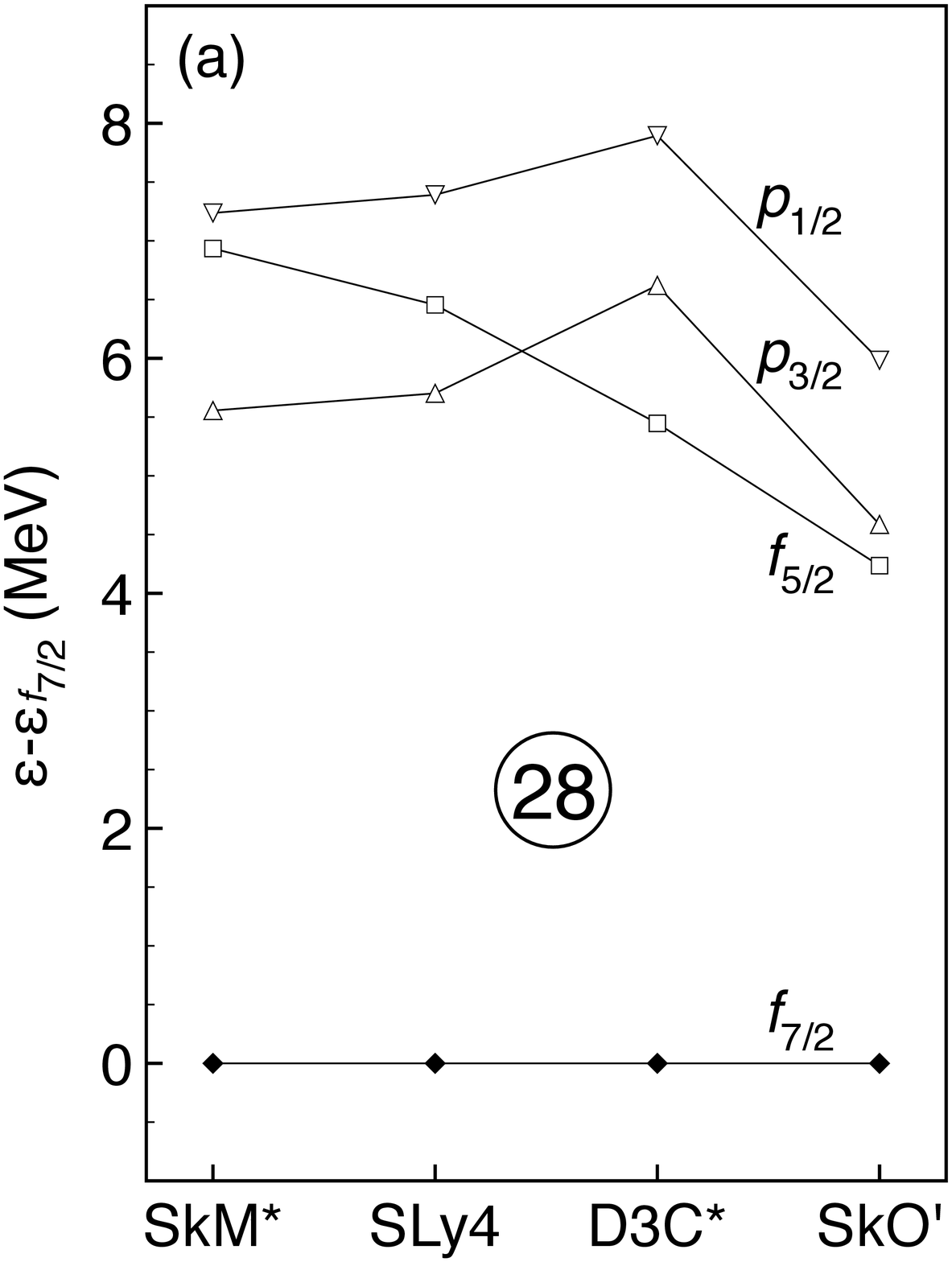}
\includegraphics[scale=0.2]{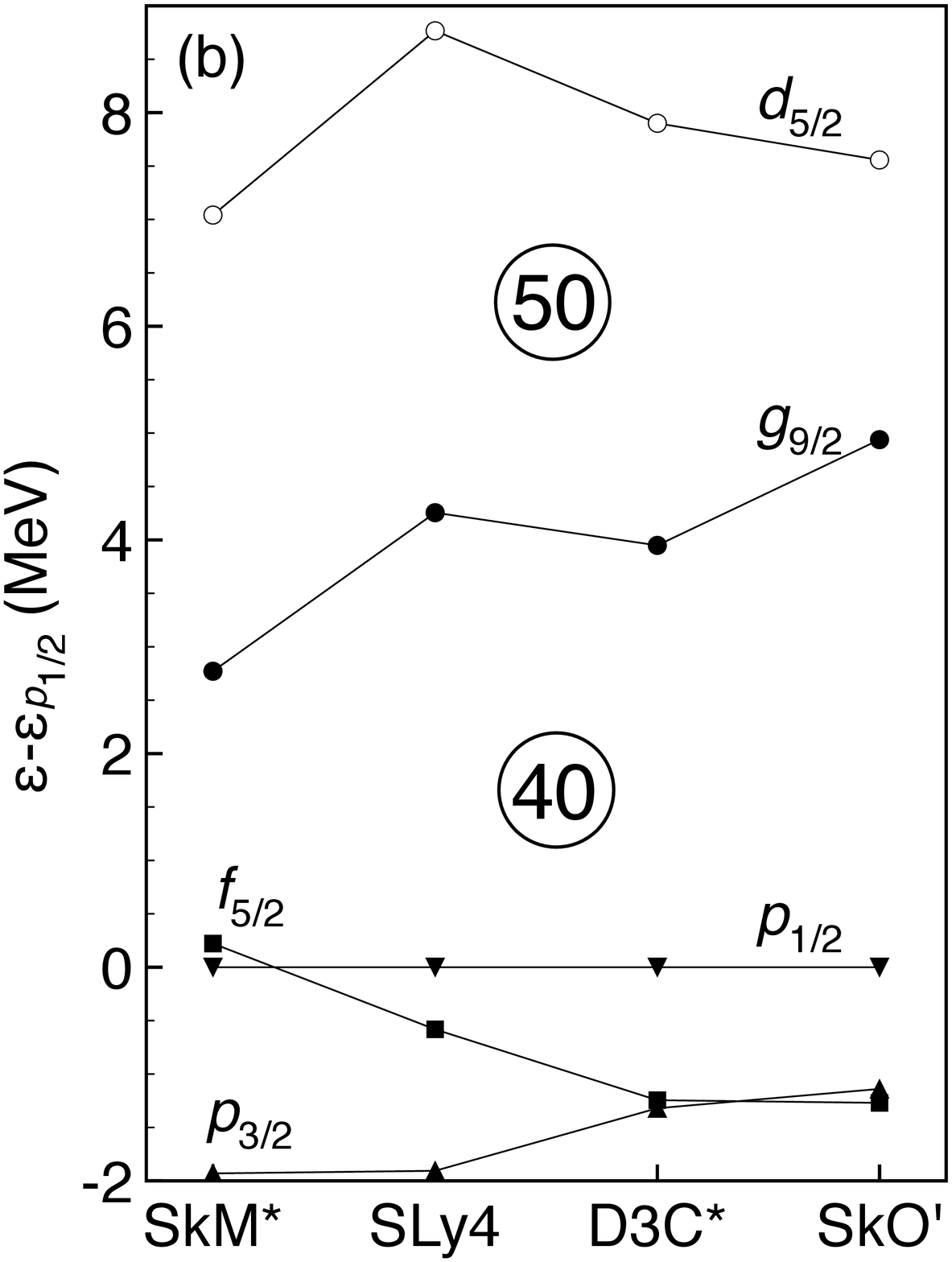}
\caption{(a) Proton's single-particle energies relative to that of the $1f_{7/2}$ orbital, and 
(b) neutron's ones relative to that of the $2p_{1/2}$ orbital calculated 
using several EDFs in $^{78}$Ni. 
The result for D3C* was obtained in Refs.~\cite{mar16,mar19}, 
and that for SkO' was obtained using {\sc HFBRAD}~\cite{ben05}.
}
\label{sp_level}
\end{center}
\end{figure}

The low-lying states relevant to the $\beta$-decay rate 
are weakly collective in the present calculation for the Ni isotopes. 
Thus, I can discuss qualitatively 
the competitive roles played by the allowed and FF transitions 
in terms of the single-particle levels around the Fermi levels. 
I show in Fig.~\ref{sp_level} the single-particle energies of neutrons and protons 
relative to that of the $\nu 2 p_{1/2}$ and $\pi 1f_{7/2}$
orbitals, respectively calculated using the SkM*, SLy4, SkO', 
and D3C* functionals.  
As discussed above, the $\nu g_{9/2} \to \pi f_{5/2}$ excitation 
is only available up to $N=50$ for the FF transition, 
and the $\nu p_{1/2} \to \pi p_{3/2}$ excitation dominantly 
contributes to the allowed transitions. 
So, the size of $N=40$ gap and the relative location 
between the $\pi f_{5/2}$ and $\pi p_{3/2}$ levels 
govern the $\beta$-decay rate for $N \leq 50$.
As the shell gap at $N=40$ is small, the energy difference between 
the allowed and FF transitions is small.
Thus, the FF contribution is small. 
In the calculation using SkM*, since the $\pi f_{5/2}$ level is located above 
the $\pi p_{3/2}$ level by 1.5 MeV, the FF contribution is strongly suppressed.  
On the other hand, as seen in the calculations with the SLy4, D3C*, and SkO' functionals, 
the FF contribution is larger as the $N=40$ gap increases.

Beyond $N=50$, the multiple FF transitions involving 
the $\nu 2d_{5/2}$ orbital are possible to occur. 
The larger the sum of the gap energy of the $N=40$ and 50, 
the larger the FF contribution is. This is because the 
energy difference between the allowed and FF excitations is large. 
For a sudden onset of the FF contribution above $N=50$, 
preferable is the situation in which 
an $N=50$ gap is large and an $N=40$ gap is instead  small.

\section{Summary}\label{summary}

I have carried out a systematic calculation of 
the $\beta$-decay rates for the neutron-rich Ni isotopes around $N=50$ 
by means of the fully selfconsistent proton-neutron-QRPA with the Skyrme EDFs. 
The experimentally observed sudden shortening of the half-lives beyond $N=50$ 
was reproduced well 
by the calculations employing the Skyrme SkM* and SLy4 functionals. 
I found that the onset of the first-forbidden (FF) transitions 
plays a decisive role for the sudden shortening of the half-life in $^{80}$Ni. 
This is due to a small sub-shell gap at $N=40$ 
and a large shell gap at $N=50$; 
the former suppresses the contribution 
from the FF transitions below $N=50$ and the latter enhances it above $N=50$. 
The present study reveals that a microscopic calculation taking 
the shell structure in neutron-rich nuclei and its associated effects 
on the FF transitions taken into account is necessary for predicting the $\beta$-decay 
rates of nuclei far from stability. 

\begin{acknowledgments} 
The author thanks J.~Engel and T.~Marketin for valuable communications, 
and N.~Van Giai for discussions. 
This work was supported by the JSPS KAKENHI (Grant No. 16K17687), and 
the JSPS-NSFC Bilateral Program for Joint Research Project on ``Nuclear mass 
and life for unraveling mysteries of the r-process".
The numerical calculations were performed on CRAY XC40 
at the Yukawa Institute for Theoretical Physics, Kyoto University, and 
on COMA (PACS-IX) at the Center for Computational Sciences, University of Tsukuba.
\end{acknowledgments}

\appendix
\section{First-forbidden $\beta$-decay rates}\label{app_FF}
The shape factor for the FF transitions is energy dependent and given as~\cite{sch66}
\begin{equation}
C(W)=k + ka W + kb/W + kc W^2,
\end{equation}
where the coefficients $k, ka, kb$, and $kc$ depend on the nuclear matrix elements and  
the maximum electron energy $W_0$. 
The non-vanishing coefficients are
\begin{align}
k&=\zeta_0^2+\dfrac{1}{9}w^2, \notag \\
kb&=-\dfrac{2}{3}\mu_1 \gamma_1 \zeta_0 W 
\end{align}
for rank 0, 
\begin{align}
k=&\zeta_1^2 +\dfrac{1}{9}(x+u)^2-\dfrac{4}{9}\mu_1\gamma_1 u (x+u) \notag\\
&+\dfrac{1}{18}W_0^2(2x+u)^2 - \dfrac{1}{18}\lambda_2 (2x-u)^2, \notag \\
ka=&-\dfrac{4}{3}uY-\dfrac{1}{9}W_0(4x^2+5u^2), \notag \\
kb=&\dfrac{2}{3}\mu_1 \gamma_1 \zeta_1(x+u), \notag \\
kc=&\dfrac{1}{18}[8u^2+(2x+u)^2+\lambda_2(2x-u)^2]
\end{align}
for rank 1, and 
\begin{align}
k&=\dfrac{1}{12}z^2(W_0-\lambda_2), \notag \\
ka&=-\dfrac{1}{6}z^2 W_0, \notag \\
kc&=\dfrac{1}{12}z^2(1+\lambda_2)
\end{align}
for rank 2, respectively with $V, Y, \zeta_0$, and $\zeta_1$ being defined by
\begin{align}
V&=\xi^\prime v+\xi w^\prime, \notag \\
\zeta_0&=V+\dfrac{1}{3}wW_0, \\
Y&=\xi^\prime y-\xi(u^\prime+x^\prime), \notag \\
\zeta_1&=Y+\dfrac{1}{3}(u-x)W_0.
\end{align}
Here, the dimensionless parameter $\xi$ is defined as $\xi=\alpha Z/2R$. 
The Coulomb functions $\mu_1$ and $\lambda_2$ are defined in terms of electron wave functions and 
depend on the electron momentum~\cite{sch66}. These values are close to unity, so 
I used the approximations $\mu_1=1$ and $\lambda_2=1$ 
as usually adopted~\cite{war88}. 

The matrix elements are related to the form factors $^{A(V)}\!F_{Jls}$ as~\cite{beh82}
\begin{subequations}
\begin{align}
w&=-R\: ^A\!F_{011}= \sqrt{3}\lambda\langle ^A\!\hat{O}^{01K}_- \rangle, \\
x&=-\dfrac{1}{\sqrt{3}}R\: ^V\!F_{110}=-\langle ^V\!\hat{O}^{11K}_-\rangle, \\
u&=-\sqrt{\dfrac{2}{3}}R\: ^A\!F_{111}=\sqrt{2}\lambda\langle ^A\!\hat{O}^{11K}_-\rangle, \\
z&=\dfrac{2}{\sqrt{3}}R\: ^A\!F_{211}=-2\lambda\langle ^A\!\hat{O}^{21K}_-\rangle, \\
\xi^\prime v &= ^A\!\!F_{000}=-\sqrt{3}\lambda\langle ^A\!\hat{O}^{000}_-\rangle, \\
\xi^\prime y&= ^V\!\!F_{101}=-\langle ^V\!\hat{O}^{10K}_-\rangle,
\end{align}
\end{subequations}
where $\lambda=-(g_A/g_V)=1.27$. 
For the FF transitions, no quenching factors were introduced for simplicity in the present calculation 
as in Ref.~\cite{mus16}. 
The primed matrix elements $w^\prime, x^\prime$, and $u^\prime$ are calculated with the operator 
multiplied by
\begin{equation}
\begin{array}{ll}
1-\dfrac{1}{5}\left( \dfrac{r}{R}\right)^2, & 0 < r < R, \\
\dfrac{R}{r}-\dfrac{1}{5}\left( \dfrac{r}{R}\right)^3, & r>R,
\end{array}
\end{equation}
where $r=\sqrt{\rho^2 + z^2}$ in the cylindrical coordinates.

\end{document}